\documentclass[prl,twocolumn,showpacs,preprintnumbers,amsmath,amssymb]{revtex4-1}

\usepackage{graphicx}
\usepackage{dcolumn}
\usepackage{bm}
\usepackage{amsmath}
\usepackage{latexsym}
\usepackage{hyperref}

\newcommand{\rfig}[1]{Fig.~\ref{#1}}

\newcommand{\be}{\begin{equation}}
\newcommand{\ee}{\end{equation}}
\newcommand{\req}[1]{Eq.~(\ref{#1})}

\begin{document}

\title{Thermoelectric effect in high mobility single layer epitaxial graphene}

\author{Xiaosong Wu$^{1,2}$, Yike Hu$^1$, Ming Ruan$^1$, Nerasoa K. Madiomanana$^1$, Claire Berger$^{1,3}$, Walt A. de Heer$^{1}$}
\affiliation{
$^1$School of Physics, Georgia Institute of Technology, Atlanta, GA 30332 \\
$^2$State Key Laboratory for Artificial Microstructure and Mesoscopic Physics, Peking University, Beijing 100871, P. R. China\\
$^3$CNRS - Institut Neel, BP 166, 38042 Grenoble cedex6, France}


\begin{abstract}
The thermoelectric response of high mobility single layer epitaxial graphene on silicon carbide substrates as a function of temperature and magnetic field have been investigated. For the temperature dependence of the thermopower, a strong deviation from the Mott relation has been observed even when the carrier density is high, which reflects the importance of the screening effect. In the quantum Hall regime, the amplitude of the thermopower peaks is lower than a quantum value predicted by theories, despite the high mobility of the sample. A systematic reduction of the amplitude with decreasing temperature suggests that the suppression of the thermopower is intrinsic to Dirac electrons in graphene.
\end{abstract}


\maketitle

Graphene, a single layer of graphite, has a unique band structure, in which electrons are described by the relativistic Dirac equation. Extensive electrical transport studies have been performed to understand Dirac electrons in the material. Compared with electrical transport, the thermoelectric properties provide complementary information to the electronic structure and the detail of electron scattering, but investigations have been started only recently \cite{Peres2007,Foster2008,Foster2009,Hwang2009,Yan2009,Zuev2009,Wei2009,Checkelsky2009,Zhu2010a,Bergman2010,Zhu2010,Yan2010}. The newly discovered topological insulators, whose electrons in surface states are also Dirac electrons, are extraordinary thermoelectric materials. The possibility of further improving the performance in its nanostructures by exploiting the Dirac nature of electrons also calls for studies on the thermoelectric response of Dirac electrons \cite{Ghaemi2010}. The thermoelectric effect of Dirac electrons has been experimentally investigated in exfoliated graphene on SiO$_2$ \cite{Zuev2009,Wei2009,Checkelsky2009}. It was found that the Mott relation, which is used to describe the thermoelectric effect in conventional 2-dimensional electron gases, is basically obeyed, but not in the vicinity of the charge neutrality point. In the quantum Hall regime, though theories predict a quantized value for the thermopower \cite{Bergman2010,Zhu2010}, experiments saw a smaller value \cite{Zuev2009,Checkelsky2009}. In those studies, the mobilities of the samples are low, in the order of a few thousand cm$^2$/V$\cdot$s.  Thus, questions have been raised about whether the quantization of the thermopower is intrinsic to graphene and can be realized in high mobility samples  \cite{Zuev2009,Checkelsky2009}. Generally, high mobility is crucial for studying the intrinsic properties of Dirac electrons. Therefore, achieving high mobility in graphene samples has been a main effort in many experiments. A few milestone experiments are indeed consequences of improved or new techniques for obtaining high mobilities, e.g. recent success in greatly improving the mobility of graphene by suspending and {\it in situ} annealing it directly led to observation of the fractional quantum Hall effect \cite{Bolotin2009,Du2009}.

In this study, we grow high mobility single layer epitaxial graphene on the silicon carbide substrate and study its thermoelectric response as a function of temperature and magnetic field. The temperature dependence deviates from the Mott relation, revealing the importance of the screening effect. In a quantized magnetic field, the thermopower is suppressed and shows unexpected temperature dependence, inconsistent with theories on the thermoelectric effect for Dirac electrons.

\begin{figure}[htb]
\includegraphics[width=0.45\textwidth]{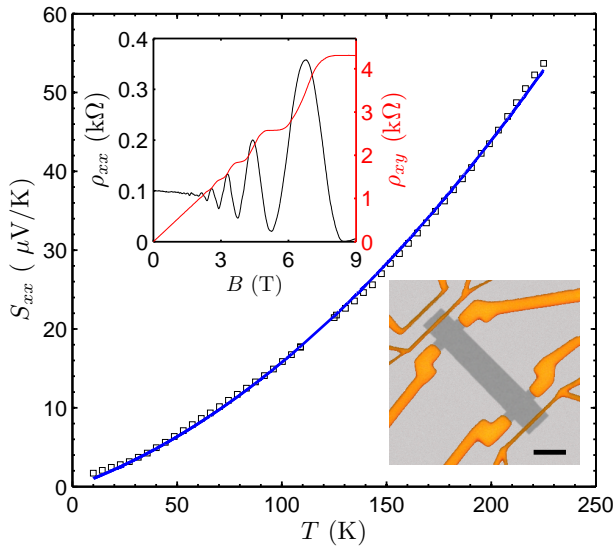}
\caption{\label{fig:SxxT} The thermopower $S_{xx}$ as a function of temperature. Open black square, experiment, solid blue line is a least square fit to $AT+BT^2$. Top left inset, the longitudinal resistivity and Hall resistance exhibit quantum Hall effect. Bottom right inset, an optical image of a graphene device for thermoelectric measurements. The scale bar is 2 $\mu$m.}
\end{figure}

High mobility single layer epitaxial graphene samples are grown on the carbon face of SiC (000$\overline{1}$) in a high vacuum furnace \cite{Berger2004,deHeer2007,Hass2008b,Wu2009}. Thermoelectric measurements are carried out following a technique developed by Kim {\it et al.} \cite{Small2003} and also used in previous thermoelectric studies on exfoliated graphene. In this technique, a local heater made of a metal line produces a temperature difference $\Delta T$ between the two ends of a sample, which gives rise to a thermoelectric voltage $\Delta V$. The temperatures of two ends are measured by two local thermometers which are also made of metal lines, as seen in the bottom right inset of \rfig{fig:SxxT}. The thermopower $S_{xx}=-\Delta V/\Delta T$. When a magnetic field is applied perpendicular to the graphene plane, a transverse thermoelectric voltage is generated, so called Nernst effect. It is defined as $S_{yx}=E_y/\nabla T$. These two determine the thermopower tensor $\boldsymbol{S}$. The thermoelectrical conductivity tensor can then be computed by $\boldsymbol{\alpha}=\boldsymbol{\sigma}\cdot \boldsymbol{S}$, where $\boldsymbol{\sigma}$ is the electrical conductivity tensor.

In our experiment, a low frequency $\omega$ ac current is applied to the heater. The voltage across the sample is then measured at 2$\omega$ frequency.  The thermopower is compared with the one measured by a DC method to rule out spurious signals. In addition, the voltage across the sample is found to be linearly proportional to the temperature difference, which confirms its thermoelectrical origin. Throughout the measurement, the temperature difference is always maintained at a level much less than the substrate temperature. 

The sample shown here is hole doped at a level of $1\times10^{12}$ cm$^{-2}$. The mobility is about $20,000$ cm$^2$/V$\cdot$s at 4 K, significantly higher than those used in previous studies. The magnetoresistance exhibits clean Shubnikov-de Haas oscillations, discernible in magnetic field down to less than 2 T, indicating its high mobility. The longitudinal resistance disappears at about 8.5 T and at the same time a plateau in the Hall resistance emerges, indicating a fully developed  quantum Hall effect, see the top left inset of \rfig{fig:SxxT}.

The Mott relation, $S_{xx}=-\dfrac{\pi^2k_B^2T}{3|e|}\dfrac{\partial \ln G} {\partial E} \Big|_{E=E_F} $, has been used to describe the diffusion thermopower for conventional electron gases. It is interesting to see whether it holds for Dirac eletrons. The temperature dependence of the thermopower $S_{xx}$ of the same is plotted in \rfig{fig:SxxT}, which displays a strong nonlinearity. It can be well fit when considering a linear dependence plus a quadratic correction, $AT+BT^2$. A dominant linear dependence was observed in exfoliated graphene, while only a slight deviation appeared at higher temperatures \cite{Zuev2009,Wei2009}. The linearity can be explained by the Mott relation. The question is what causes the deviation. Phonon drag effect is unlikely, not only because the electron-phonon coupling is weak in graphene, but because it gives rise to a large power index, usually no less than 4 \cite{[test pre][test apd]Fletcher1999}. Hwang {\it et al.} studied the temperature dependence of the thermopower of graphene \cite{Hwang2009}. They found that when the screening effect and its temperature dependence are taken into account, a quadratic correction to the thermopower appears. Note that the dielectric constant of SiC is about 10, a factor of over 2 higher than that of SiO$_2$. The screening is consequently stronger in epitaxial graphene than in exfoliated graphene on SiO$_2$, which accounts for the stronger nonlinearity observed here. 

\begin{figure}[htb]
\includegraphics[width=0.45\textwidth]{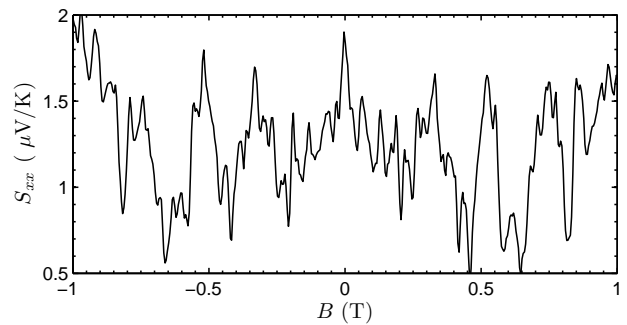}
\caption{\label{fig:USFs} The thermopower in low magnetic field displays strong and reproducible fluctuations.}
\end{figure}

The magnetic field dependence of the thermopower $S_{xx}$ at low temperature displays reproducible fluctuations in low field, as shown in \rfig{fig:USFs}. The fluctuations are manifest of phase coherent transport, the same origin of the universal conductance fluctuations (UCFs) \cite{Beenakker1991}. The large amplitude of the oscillations indicates a long coherence length, which is consistent with the high mobility of the sample.

\begin{figure}[htb]
\includegraphics[width=0.45\textwidth]{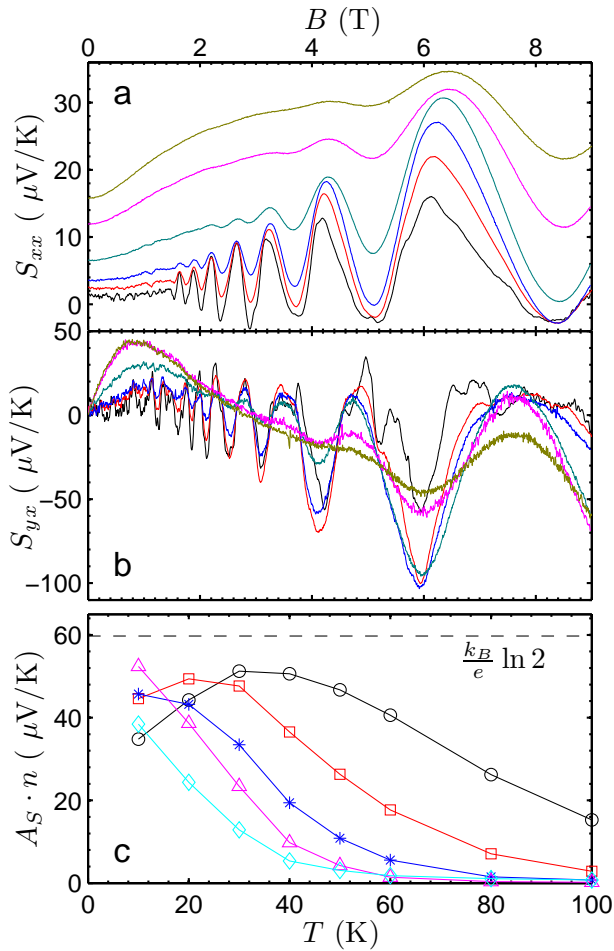}
\caption{\label{fig:SB} The magnetic field dependence of the thermopower $S_{xx}$ (panel {\bf a}) and the Nernst signal $S_{yx}$ (panel {\bf b}) are plotted for $T=$10, 20, 30, 50, 80, 100 K, denoted by black, red, blue, dark green, magenta and dark yellow, respectively. Both show oscillations that are periodic in inverse magnetic field. The amplitudes of the oscillations $A_S$ are multiplied by the Landau index $n$ and plotted as a function of temperature in panel {\bf c} for $n=2, 3, 4, 5, 6$ (circle, square, star, triangle and diamond). The dashed line marks $\dfrac{k_B}{e}\ln 2$.}
\end{figure}

Girvin and Jonson have calculated the thermoelectric tensor for a conventional 2-dimensional electron gas subject to a quantized magnetic field using a generalized Mott formula \cite{Girvin1982,Jonson1984}. They found that the thermopower exhibits a large peak when a Landau level is half filled. The peak value is quantized at $\dfrac{k_B}{e}\dfrac{\ln 2}{n+1/2}$ and independent of temperature and magnetic field. Here $k_B$, $e$ and $n$ denote the Boltzmann constant, the electron charge and the Landau index, respectively. Recent theoretical work confirmed the same quantization for Dirac electrons in graphene, except $n+1/2$ is replaced by $n$ because of the anomalous Berry's phase of $\pi$ \cite{Bergman2010,Zhu2010}. We have measured $S_{xx}$ and $S_{yx}$ in high magnetic field. Both show strong quantum oscillations periodic in $1/B$, as seen in \rfig{fig:SB}{\bf a} and {\bf b}. The oscillations are the consequences of formation of Landau levels and can be understood by the change of the density of state at the Fermi level. That is, when the Fermi level lies in the localized states, $S_{xx}$ becomes zero because of absence of diffusion. The oscillations of $S_{xx}$ slightly overshoot $y=0$, resulting in small negative minima of about -2 $\mu$V/K. The amplitude of $S_{xx}$ oscillations at each Landau level $n$ is obtained by subtracting the minima contour from the corresponding peak height. In the panel {\bf c} of \rfig{fig:SB}, the amplitude $A_S$ multiplied by its Landau index $n$ is plotted against temperature. $A_S\cdot n$ for all Landau levels increases with decreasing temperature at high temperature, which can be explained by reduction of thermal broadening of Landau levels. At low temperature in the quantum Hall regime, according to the theoretical result, $A_S\cdot n$ will collapse to a temperature independent value of $\dfrac{k_B}{e}\ln 2\sim 59.6 \mu$V/K. Interestingly, in our experiment, the $A_S\cdot n$ reaches 52 $\mu$V/K and then starts to decrease, showing a turning point. The lower the Landau index $n$, the higher temperature the turning point occurs at and the stronger the suppression is at 10 K. The systematic trend suggests that deep in the quantum Hall regime, $S_{xx}$ is suppressed, contradicting with a temperature independent quantum value predicted by theories.

On the other hand, $S_{yx}$ oscillates around zero. The phase of the $S_{yx}$ oscillations shifts by about $\pi$/4 with repect to $S_{xx}$, consistent with the generalized Mott formula. Note that, on top of the oscillations, $S_{yx}$ also exhibits fluctuations, which are stronger than those in $S_{xx}$. This is due to the shorter distance between two Hall probes that are used to measure $S_{yx}$. The slope of $S_{yx}$ in low field is linear in temperature. The Nernst effect under weak magnetic field in graphene has been studied theoretically \cite{Yan2010}. It was found that the linearity depends on the details of electron scattering. It would be interesting to extract such information from our experiment. Unfortunately, the theoretical result is numerical and direct comparison between the theory and our experiment cannot be easily made.

Another feature is that $S_{yx}$ changes sign at 2.8 T even when the temperature is so high that Landau levels are thermally smeared out. The question may be raised whether two types of carriers coexist in the sample. However, a perfectly straight Hall resistance as a function of magnetic field, seen in the inset of \rfig{fig:SxxT}, rules out this possibility. Moreover, no clear indication of two frequencies can be seen in the SdH oscillations. Unlike the resistivity, which is determined by the scattering time, the thermoelectric effect is sensitive to the energy dependence of the scattering time. Therefore, the intriguing features seen in the low field $S_{yx}$, which are in sharp contrast to the trivial behavior of the Hall resistance, are most likely manifest of the detail of scattering mechanism. Further investigation may provide insight on the scattering processes in graphene.

\begin{figure}[htb]
\includegraphics[width=0.45\textwidth]{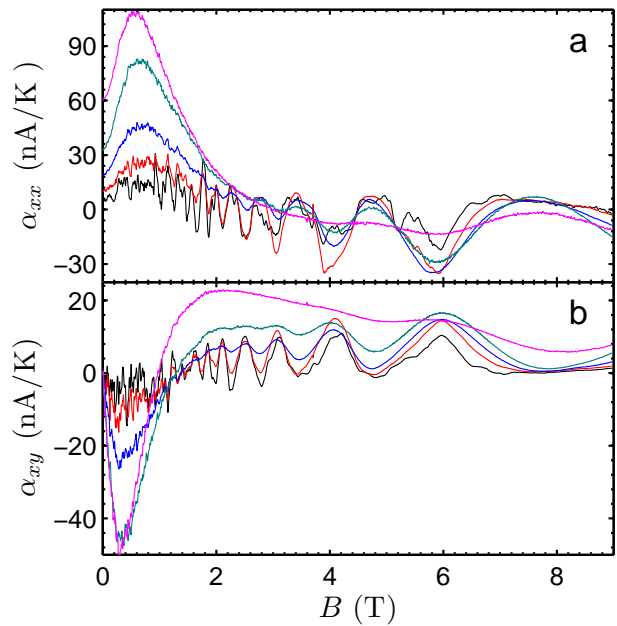}
\caption{\label{fig:AlphaB} The magnetic field dependence of the longitudinal thermoelectric conductivity $\alpha_{xx}$ and transverse thermoelectric conductivity $\alpha_{xy}$ at $T=$10, 20, 40, 60, 100 K (black, red, blue, dark green and magenta), panel {\bf a} and panel {\bf b}, respectively.}
\end{figure}

In some cases, it is intuitive to look at the thermoelectrical conductivity $\boldsymbol{\alpha}$. For instance, $\alpha_{xy}$ can be linked to electron entropy \cite{Bergman2010}. Having measured the conductivity tensor $\sigma$ and the thermopower tensor $S$, the calculation of $\alpha$ is straightforward, $\boldsymbol{\alpha}=\boldsymbol{\sigma}\cdot \boldsymbol{S}$. In particular:
\begin{align}
\alpha_{xx}=\sigma_{xx}S_{xx}+\sigma_{xy}S_{yx} \label{alphaxx} \\
\alpha_{xy}=\sigma_{xy}S_{xx}-\sigma_{xx}S_{yx} \label{alphaxy}
\end{align}
In \rfig{fig:AlphaB}, we plot the two components of the tensor $\alpha_{xx}$ and $\alpha_{xy}$ as a function of magnetic field at different temperatures. Similar to $S_{yx}$, both components display quantum oscillations at low temperatures and a change of the sign in the intermediate field regime at high temperatures.

Like $S_{xx}$, $\alpha_{xy}$ has also been predicted to quantize at $\dfrac{gk_Be}{h}\ln 2 \sim 9.2$ nA/K, when a Landau level is half filled. Here $g$ is the total degenercy. Attempts to experimentally test both quantization have been made, while no link between them has been provided \cite{Zuev2009,Checkelsky2009}. The formulae for two quantum values are very similar. In fact, they are closely related, as explained in the following manner. Note that the oscillations of $S_{xx}$ is in phase with $\sigma_{xx}$ and ahead of $S_{xy}$ by $\pi/4$. When a Landau level is half filled, $S_{yx}$ becomes zero, hence the second term on the right side of \req{alphaxy}. At the same time, $\sigma_{xy}$ is at the middle of two quantum Hall plateaus. Therefore, according to \req{alphaxy}, $\alpha_{xy}=\sigma_{xy}S_{xx}=\dfrac{gne^2}{h}\cdot \dfrac{k_B}{e}\dfrac{\ln 2}{n}=\dfrac{gk_Be}{h}\ln 2$. It is clear that a suppression of $S_{xx}$ peak will lead to a suppression of $\alpha_{xy}$. However, in our experiment, the amplitude of $\alpha_{xy}$ oscillations reaches 15 nA/K, over 50\% larger than the quantum value 9.2 nA/K, inconsistent with the smaller value of $S_{xx}$. We want to point out that $S_{yx}$ is subject to a large uncertainty because the temperature gradient can not be measured directly, instead extrapolated, not to mention that $\alpha_{xy}$ is calculated from four experimentally measured quantities. Consequently, we believe that it is much more reliable to test the quantization of $S_{xx}$ than $\alpha_{xy}$. Nevertheless, a similar temperature dependence of $\alpha_{xy}$ as that of $S_{xx}$ at low temperature are observed.

Most of the theories on the thermoelectric effect in graphene have found that the Mott relation holds well except for very low carrier density and $S_{xx}$ in the quantum Hall regime has maxima of $\dfrac{k_B}{e}\dfrac{\ln 2}{n}$. Previous experiments all seem more or less consistent with these predictions. However, it is worth noting that a $S_{xx}$ maximum smaller than the theoretical prediction was always observed in experiments and ascribed to low mobilities of the samples \cite{Zuev2009,Checkelsky2009}. In our sample, which has a significantly higher mobility, a suppression of $S_{xx}$ is still seen. This suggests that the suppression of the $S_{xx}$ peak may be intrinsic to graphene. A systematic reduction of $S_{xx}$ with decreasing temperature for different Landau levels further strongly supports the suggestion. Although most theories predict a temperature independent $S_{xx}$ peak in the QHE regime, there is one exception to our best of knowledge. Zhu {\it et al.} have studied the temperature dependence of the amplitudes of both $S_{xx}$ and $\alpha_{xy}$ peaks\cite{Zhu2010}. A linear dependence was found at low temperature, while it saturates at the quantum value at high temperature. However, the numeric computation was performed for extremely high magnetic field in the order of hundreds of Tesla. It is not clear if the suppression we saw in a much lower field ($<$ 10 Tesla) is indeed the linear dependence predicted by the theory. Bergman {\it et al.} have suggested that further theoretical work, taking into account inelastic processes, might provide a clue \cite{Bergman2010}.

In conclusion, we study the thermoelectric response of high mobility single layer epitaxial graphene. We observe a strong deviation from the Mott relation, even when the system is not in the vicinity of the Dirac point. In a magnetic field, while the Hall resistivity displays a trivial linear dependence, the Nernst signal shows a nonmonotonic behavior, which we believe is related to different scattering mechanisms. In the quantum Hall regime, contrary to theories, the amplitude of the thermopower peaks is lower than the quantum value, although the mobility of the sample is high. The suppression of the thermopower is further confirmed by its systematic reduction with decreasing temperature. To understand these behaviors, further theoretical work that takes into account inelastic processes of Dirac electrons is needed.

This work was supported by NSF grant DMR-0820382 and the W. M. Keck Foundation. X.W. thanks Xin-Zhong Yan for helpful discussions.

\end{document}